% Revised after NPB referee report, submitted to hep-th on ?

\input harvmac
\input epsf

\overfullrule=0pt
\abovedisplayskip=12pt plus 3pt minus 3pt
\belowdisplayskip=12pt plus 3pt minus 3pt
%macros
%
\def\tilde{\widetilde}
\def\bar{\overline}
\def\to{\rightarrow}

\def\tPhi{{\tilde\Phi}}
\def\cN{{\cal N}}
\def\bN{{\bar N}}
\font\zfont = cmss10 %scaled \magstep1
\font\litfont = cmr6

\def\bigone{\hbox{1\kern -.23em {\rm l}}}
\def\ZZ{\hbox{\zfont Z\kern-.4emZ}}
\def\half{{\litfont {1 \over 2}}}

% References

\lref\klebwit{I. Klebanov and E. Witten, {\it ``Superconformal Field 
Theory on Threebranes at a Calabi- Yau Singularity''}, (1998);
hep-th/9807080.}
\lref\maldacena{J. Maldacena, {\it ``The Large N Limit of
Superconformal Field Theories and Supergravity''},
Adv. Theor. Math. Phys. {\bf 2} (1998) 231; hep-th/9711200.}
\lref\gubs{S. S. Gubser, {\it ``Einstein Manifolds and Conformal Field
Theories''}, (1998); hep-th/9807164.}
\lref\gubkleb{S. S. Gubser and I. R. Klebanov, {\it ``Baryons and Domain
Walls in an $\cN =1$ Superconformal Gauge Theory''}, (1998);
hep-th/9808075.}
\lref\bsv{M. Bershadsky, V. Sadov and C. Vafa, {\it ``D-strings on
D-manifolds''}, Nucl. Phys. {\bf B463} (1996) 398; hep-th/9510225. }
\lref\stromconif{A. Strominger, {\it ``Massless Black Holes and
Conifolds in String Theory''}, Nucl. Phys. {\bf B451} (1995) 96;
hep-th/9504090.}
\lref\bho{E. Bergshoeff, C. Hull and T. Ortin,
{\it `` Duality in the Type II Superstring Effective 
Action''}, Nucl. Phys. {\bf B451} (1995) 547; hep-th/9504081. }
\lref\hanwit{A. Hanany and E. Witten, {\it ``Type IIB Superstrings, BPS
Monopoles, and Three-Dimensional Gauge Dynamics''}, Nucl. Phys. {\bf 
B492} (1997) 152; hep-th/9611230.}
\lref\witneqone{E. Witten, {\it ``Branes and the Dynamics of QCD''},
Nucl. Phys. {\bf B507} (1997) 658; hep-th/9706109. }
\lref\hoo{K. Hori, H. Ooguri and Y. Oz, {\it ``Strong Coupling
Dynamics of Four-dimensional $\cN=1$ Gauge Theories from M-theory
Fivebranes''}, Adv. Theor. Math. Phys. {\bf 1} (1997) 1;
hep-th/9706082.} 
\lref\brand{A. Brandhuber, N. Itzhaki, V. Kaplunovsky, J. Sonnenschein
and S. Yankielowicz, {\it ``Comments on the M Theory Approach to N=1
SQCD and Brane Dynamics''}, Phys.Lett. {\bf B410} (1997) 27; 
hep-th/9706127.}
\lref\gutcurio{B. Andreas, G. Curio and D. Lust, {\it ``The Neveu-Schwarz
 Five Brane and its dual Geometries''}, hep-th/9807008. }
\lref\gnt{J. P. Gauntlett, {\it ``Intersecting Branes''}, 
hep-th/9705011.}
\lref\tseytlin{A.A. Tseytlin, {\it ``Harmonic Superpositions of
M-branes''}, Nucl. Phys. {\bf B475} (1996) 149; hep-th/9604035.}
\lref\tseytrev{A.A. Tseytlin, {\it ``Composite BPS Configurations of 
$p$-branes in Ten Dimensions and Eleven Dimensions''}, 
Class. Quant. Grav. {\bf 14} (1997) 2085; hep-th/9702163.}
\lref\lenvafa{N. C. Leung and C. Vafa, {\it ``Branes and Toric Geometry''
}, Adv. Theor. Math. Phys. {\bf 2} (1998) 91; hep-th/9711013.}
\lref\candelas{P. Candelas and X. C. de la Ossa, {\it ``Comments on 
Conifolds''}, 
 Nucl. Phys. {\bf B342} (1990) 246.}
\lref\mori{D. Morrison and R. Plesser, {\it ``Non- Spherical Horizons-I''
}, (1998); hep-th/9810201.}
\lref\afhs{B.S. Acharya, J.M. Figueroa-O'Farrill, C.M. Hull and
B. Spence, {\it ``Branes at Conical Singularities and Holography''},
hep-th/9808014.}
\lref\angel{A. Uranga, {\it ``Branes configurations for Branes at
Conifolds''}, (1998); hep-th/9811004.}
\lref\gnttown{J. P. Gauntlett, G. W. Gibbons, G. Papadopoulos and P.K.
Townsend, {\it ``Hyperkahler Manifolds and Multiply Intersecting 
Branes''}, Nucl. Phys. {\bf B500} (1997) 133; hep-th/9702202.} 
\lref\papad{G. Papadopoulos and A. Teschendorff, {\it ``Multiangle 
Five-brane Intersections''}, hep-th/9806191\semi
G. Papadopoulos and A. Teschendorff, {\it ``Grassmannians,
Calibrations and Five-brane Intersections''}, hep-th/9811034.}
\lref\wkinprogress{Work in progress.}
\lref\witfourd{E. Witten, {\it ``Solutions of Four Dimensional Field 
Theories via M-Theory},  Nucl. Phys. {\bf B500} (1997) 3;
hep-th/9703166.} 
\lref\givkut{A. Giveon and D. Kutasov, {\it ``Brane Dynamics and Gauge
Theories''}, (1998); hep-th/9802067.} 
\lref\kehagias{A. Kehagias, {\it ``New Type IIB Vacua and Their
F-theory Interpretation''}, Phys. Lett. {\bf B435} (1998) 337;
hep-th/9805131.}
\lref\ahm{O. Aharony, A. Fayyazuddin and J. Maldacena, {\it ``The
Large-N Limit of $\cN=2$, $\cN=1$ Field Theories from Threebranes in
F-theory''}, JHEP {\bf 07} (1998) 013; hep-th/9806159.}  
\lref\hanzaf{A. Hanany and A. Zaffaroni, {\it ``On The Realisation of
Chiral Four Dimensional Gauge Theories using Branes''}, JHEP {\bf 05}
(1998) 001; hep-th/9801134.}
\lref\hanuranga{A. Hanany and A. Uranga, {\it ``Brane boxes and Branes on 
Singularities''}, JHEP {\bf 05} (1998) 013; hep-th/9805139.}
\lref\seiwitone{N. Seiberg and E.Witten, {\it ``Electric-Magnetic
duality, Monopole Condensation and Confinement in $\cN =2$
Supersymmetric Yang- Mills Theory''}, Nucl. Phys. {\bf B426} (1994)
19; hep-th/9407087.}
\lref\donagiwitten{R. Donagi and E. Witten, {\it ``Supersymmetric 
Yang-Mills Theory and Integrable Systems''}, Nucl. Phys. {\bf B460}
(1996) 296; hep-th/9510101.}
\lref\tong{N. Dorey and D. Tong, {\it ``The BPS Spectra of Gauge
Theories in Two and Four Dimensions''}, to appear.}
\lref\lnv{A. Lawrence, N. Nekrasov and C. Vafa, {\it ``On Conformal
Field Theories in Four Dimensions''}, Nucl. Phys. {\bf B533} (1998)
199; hep-th/9803015.}
\lref\ahahan{O. Aharony and A. Hanany, {\it ``Branes, Superpotentials
and Superconformal Fixed Points''}, Nucl. Phys. {\bf B504}
(1997) 239; hep-th/9704170.}
%\lref\ohtas{T. Kitao, K. Ohta and N. Ohta, {\it ``Three-Dimensional 
%Gauge Dynamics from Brane Configurations with (p,q)-Fivebrane''},
%Nucl. Phys. {\bf B539} (1999) 79; hep-th/9808111.} 

{\nopagenumbers
\Title{\vtop{\hbox{hep-th/9811139}
\hbox{IASSNS-HEP-98/95}
\hbox{TIFR/TH/98-43}}}
{\centerline{Brane Constructions, Conifolds and M-Theory}}
\centerline{Keshav Dasgupta\foot{E-mail: keshav@ias.edu}}
\vskip 3pt
\centerline{\it School of Natural Sciences, Institute for Advanced Study}
\centerline{\it Olden Lane, Princeton NJ 08540, U.S.A.}
\vskip 10pt
\centerline{Sunil Mukhi\foot{E-mail: mukhi@tifr.res.in}}
\vskip 3pt
\centerline{\it Tata Institute of Fundamental Research,}
\centerline{\it Homi Bhabha Rd, Mumbai 400 005, India}
\vskip 5pt

\ \smallskip
\centerline{ABSTRACT}

We show that a set of parallel 3-brane probes near a conifold
singularity can be mapped onto a configuration of intersecting branes
in type IIA string theory. The field theory on the probes can be
explicitly derived from this formulation. The intersecting-brane
metric for our model is obtained using various dualities and related
directly to the conifold metric. The M-theory limit of this model is
derived and turns out to be remarkably simple. The global symmetries
and counting of moduli are interpreted in the M-theory picture.

\Date{November 1998}
\vfill\eject}
\ftno=0

\newsec{Introduction}

Recently, a system of parallel 3-branes in the presence of a conifold
singularity has received some attention\refs\klebwit. The $\cN=1$
supersymmetric gauge theory on the 3-branes has been deduced
indirectly from properties of the conifold. For a large number $N$ of
3-branes, the system is conjectured to be dual to a certain IIB string
compactification, extending the AdS-CFT correspondence\refs\maldacena\
in an interesting way. One of the notable features of this system is
that the compact space which occurs on the string theory side of the
duality is not even locally $S^5$.

This system has been investigated further in Refs.\refs{\gubs,\gubkleb},
where baryon-like chiral operators built out of products of $N$ chiral
superfields were identifies with D3 branes wrapped over the three
cycle of the compact space $T^{1,1}$. Also a D5 brane wrapped over a
two cycle of $T^{1,1}$ was identified with a domain wall in
$AdS_5$. Upon crossing it, the gauge group is argued to change from
$SU(N)\times SU(N)$ to $SU(N)\times SU(N+1)$.

Our goal in what follows will be to derive the conformal field theory
on 3-branes at the conifold singularity using a version of the brane
construction pioneered by Hanany and Witten\refs\hanwit.  This
construction will enable us to explicitly read off the spectrum and
other properties of the conformal field theory on 3-branes at a
conifold. 

It has been argued some time ago that the conifold singularity (in the
absence of a transverse 3-brane) is dual to a system of perpendicular
NS 5-branes intersecting over a $3+1$ dimensional world
volume\refs{\bsv}. In this formulation, a 3-brane wrapped over a
3-cycle that shrinks as one approaches the conifold limit from the
``deformation'' side is replaced by an open D-string connecting the
two NS branes. By an S-duality, this can be replaced by a fundamental
open string connecting perpendicular D 5-branes, and one can now see
in perturbation theory the famous massless hypermultiplet whose
presence ``cures'' the conifold singularity\refs\stromconif. However
this picture, though it provided initial inspiration for the present
work, will not be directly useful for us.

Starting from a slightly different viewpoint, we will argue that the
conifold singularity is represented by a configuration of two type IIA
NS 5-branes that are rotated with respect to one another, and located
on a circle, with D 4-branes stretched between them from both
sides. The result is an elliptic version of a model studied in
Refs.\refs{\hoo,\witneqone,\brand}, where it was used to analyse pure
$\cN=1$ supersymmetric QCD. In particular, this clarifies the
relationship between the AdS duals to branes at quotient singularities
and branes at conifolds, explaining why the latter is a relevant
deformation of the former.

After deriving our model, we will use the supergravity solution for
intersecting branes to give a heuristic but more explicit map from
brane configurations to conifolds. This will allow us to make many
identifications more precise, and also to argue that the separation of
the NS and NS' 5-branes along one of the common transverse directions
can be interpreted as turning on a constant flux of the NS-NS
$B$-field on the type IIB side.

Finally, we investigate the M-theory limit of our
model\refs\witfourd. This has been useful in the past in studying the
solution of two kinds of nontrivial models, those with nonzero beta
function and those which are conformally invariant. Our model falls in
the latter class, but unlike its $\cN=2$ supersymmetric counterpart,
it has a surprisingly simple lift to M-theory as we will show. Some of
the continuous and discrete global symmetries of the model, and the
counting of moduli, come out naturally in the M-theory picture. In
particular the RR $B$-field will make its appearance symmetrically
with the NS-NS $B$-field, describing separation of branes along the
$x^{10}$ direction.

This construction allows several interesting generalizations,
which will not be analysed here. A class of conifold-like
singularities parametrized by two integers $(n,n')$ was investigated,
for example, in Ref.\refs\bsv. In this case we expect to find several
rotated NS 5 branes arranged around a circle, a model recently
investigated by Uranga\refs\angel. More general conical and other
singularities have been addressed from the $AdS$ viewpoint
in Ref.\refs\afhs\ and more recently in Ref.\refs\mori.

\newsec{Conifolds and Intersecting Branes}

Let us review the basic idea in Ref.\refs\bsv\ to map a conifold or
generalization thereof to a set of intersecting NS and NS'
5-branes. The equation of a conifold,
\eqn\conifold{
(z_1)^2 + (z_2)^2 + (z_3)^2 + (z_4)^2 = 0 }
can be rewritten
\eqn\rewrit{
\eqalign{
(z_1)^2 + (z_2)^2 &= \zeta \cr
(z_3)^2 + (z_4)^2 &= -\zeta}}
which describes two degenerating tori varying over a $P^1$ base. By
performing two T-dualities, one over a cycle of each torus, one ends
up with a pair of NS 5-branes which are locally along the directions
$x^1, x^2, x^3, x^4, x^5$ and $x^1, x^2, x^3, x^8, x^9$ respectively,
where $x^4$ and $x^8$ are the directions along which T-duality was
performed. The D3-brane that shrinks at the conifold singularity
becomes a D-string stretching between these NS 5-branes in this
language.

If we place $N$ D3-branes transverse to the conifold singularity
(i.e., along the $(x^1,x^2,x^3)$ directions), then after these dualities
they turn into D5-branes covering the 2-torus along the $4-6$
directions. The result is identical to a ``brane
box''\refs{\hanzaf,\hanuranga}, but unfortunately this does not seem
to be a useful description of the model in which we are interested.

According to Ref.\refs\hanuranga, such a model should have gauge group
$U(N)$ and $\cN=4$ supersymmetry, unless the brane box is
``twisted''. In the latter case it would have the desired gauge group
$U(N)\times U(N)$ but still $\cN=2$ supersymmetry rather than $\cN=1$
which we expect. We will return to this point in a subsequent
section. Presumably this model is not actually incorrect, but rather
the standard techniques to analyse brane configurations are less
useful in this description. The model that we construct in the next
section will turn out to be easy to analyse and to describe all
qualititative features of branes at conifolds rather well.

\newsec{Branes at Conifolds and Fibred Brane Configurations}

Let us write Eq.\conifold\ above as
\eqn\conifagain{
(z_1)^2 + (z_2)^2 + (z_3)^2 = - (z_4)^2}
In this form it describes the $Z_2$ ALE space $R^4/Z_2$ blown up by a
$P^1$ of size $|z_4|$. Thus it can be thought of as a fibration where
the base is the $z_4$ plane and the fibre is an ALE (Eguchi-Hanson)
space of linearly varying scale size.

Let us choose conventions in which $z_4 = x^4 + ix^5$ and the
directions $x^6,x^7,x^8,x^9$ describe the ALE space embedded in
$z_1,z_2,z_3$. The ALE space is centred at
$(x^6,x^7,x^8,x^9)=(0,0,0,0)$. Moreover, at $(x^4,x^5)=(0,0)$ the
scale size shrinks to zero and there is a singularity, the node of the
conifold.

Suppose we are very close to this singularity. Then the ALE space can
be replaced with a positive-mass two-centre Taub-NUT space. The scale
size of the ALE space is traded for the distance separating the two
centres in the Taub-NUT space, hence this distance also varies
linearly as a function of $x^4,x^5$.

This means that we have a pair of 5-brane Kaluza-Klein monopoles
filling the $x^1,x^2,x^3$ directions and separating from each other
linearly along $x^6, x^7,x^8,x^9$ as a function of $x^4,x^5$. This
function must be holomorphic in suitable complex coordinates in order
to produce a supersymmetric model. Hence we can choose the two KK
5-branes to lie at $(x^6,x^7,x^8,x^9)=\alpha (0,0,x^4,x^5)$. As a
result, they intersect over a 3-brane in the $x^1,x^2,x^3$ directions,
at the point $x^4=x^5=0$. Thus we will replace the conifold by this
configuration of intersecting KK monopoles.

A T-duality along the $x^6$ direction converts these KK monopoles to a
pair of NS 5-branes aligned in the same way.  More generally, we can
separate the two NS5-branes along the $x^6$ and $x^7$ directions.  The
$x^6$ coordinates of these branes are actually determined by the
background $B$-field if there is one. Turning on such a $B$-field
causes the $x^6$ separation to be proportional to the integral of the
$B$-field over the (vanishing) 2-cycle of the original ALE
space. Hence the branes are also separated in the $x^6$ direction if
the $B$ field is nonzero. This will be confirmed in a subsequent
section.

Now, if $N$ D3-branes are placed transverse to the conifold then upon
performing the T-duality described above, they turn into D4-branes
stretched along the compact $x^6$ direction. These 4-branes are
``broken'' twice, once on each NS 5-brane, so there are really two
independent segments for each 4-brane on the $x^6$ circle.

Let us now look at the limit in which the number $N$ of D3-branes
(which are now D4-branes) becomes large. For parallel NS5-branes, the
spacetime becomes $AdS_5\times S^5/Z_2$ with a total of 16
supersymmetries. For our rotated branes, the transverse space is
similar except that the singular circle on $S^5$ is blown up by a
$P^1$. This $P^1$ is in fact present all over the $4,5$ plane, but its
size is varying with distance away from the centre. However, since the
fixed circle of $S^5/Z_2$ arises when
\eqn\fixedcirc{
(x^6,x^7,x^8,x^9)\rightarrow - (x^6,x^7,x^8,x^9)}
it lies at the origin of $6,7,8,9$ and along a circle in the $4,5$
plane. Along this circle, the $P^1$ has a constant size.

It has been noted that precisely this blowup of $S^5/Z_2$ gives rise
to the smooth Einstein manifold $T_{1,1}$. This blowup has some
unusual properties relative to conventional blowups of ALE spaces: (i)
it is a relevant and not a marginal deformation in the brane field
theory, (ii) it breaks 16 supersymmetries down to 8. At this stage we
can see roughly how these arise. The $P^1$ actually varies in size
over the full spacetime, hence one may expect that it is not just a
marginal perturbation, even though it is a constant-size blowup of the
fixed locus in $S^5/Z_2$. The branes are not parallel, but rotated in
a definite way. This makes the adjoint fields massive, inducing
precisely the correct relevant perturbation to break $\cN=2$
supersymmetry to $\cN=1$. We will see that the induced mass terms in the
superpotential are antisymmetric under exchange of gauge groups, as
expected from Ref.\refs\klebwit.

\newsec{Analysis of the Model}

In order to draw a figure of the model that we will be discussing, it
is convenient to suppress certain directions. The $x^1,x^2,x^3$
directions are always suppressed as they correspond to the noncompact
dimensions in which the field theory lives. It is convenient also to
think of the coordinates $x^4+ix^5$ and $x^8+ix^9$ as representing one
complex dimension each. We also suppress the $x^7$ direction. The
configuration relevant to the conifold is then as in Fig. 1.
\bigskip

\centerline{\epsfbox{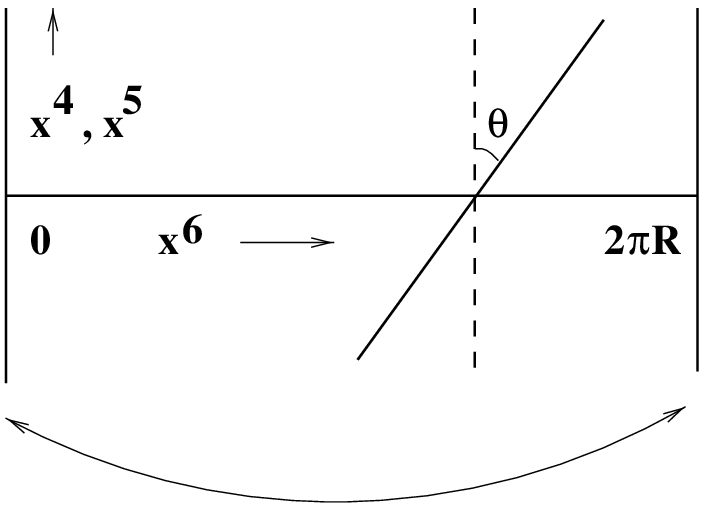}}\nobreak
\centerline{Fig. 1: Brane configuration for the conifold theory}
\bigskip

The gauge group of this configuration is straightforward to read
off. To start off, we had $N$ D3-branes transverse to the conifold. By
T-duality these have turned into $N$ D4-branes, and moreover they
stretch from the first NS5-brane to the second and back around the
$x^6$ circle to the first. Thus they have ``broken'' into two
segments, and from standard arguments we expect one $U(N)$ gauge group
from each segment. This naive $U(N)\times U(N)$ gauge group contains
two $U(1)$ factors, one of which is the pure centre of mass motion and
the other decouples by standard reasoning as in
Ref.\refs\witfourd. Thus the gauge group is $SU(N)\times SU(N)\times
U(1)$, exactly as desired.

Next let us look for the matter multiplets. Open strings stretching
across the two points where the D4-brane is split by the NS5-branes
correspond as usual to bi-fundamentals. In the language of $\cN=2$
supersymmetry we get two bi-fundamental hypermultiplets, which
decompose under $\cN=1$ supersymmetry as two chiral multiplets
$A_1,A_2$ in the $(N,\bN)$ of $SU(N)\times SU(N)$ and two more chiral
multiplets $B_1,B_2$ in the $(\bN,N)$.

This is precisely the postulated field content of the model of $N$
D3-branes at a conifold. However, so far we have just reproduced
fields which arose already in the $\cN=2$ model. One difference now
becomes apparent: there are no moduli for the centre of mass of the
D4-branes to move in the $4,5$ or $8,9$ directions. This is well-known
to imply that the adjoint has acquired a mass. Moreover, there is no
way to separate the $N$ D4-branes from each other and go to the
Coulomb branch of $SU(N)\times SU(N)$. But this is exactly what we
expect from the analysis of Ref.\klebwit.

Moreover, because of their immobility, the D4-branes on opposite sides
of each NS5-brane cannot split off from each other. This means that
the bi-fundamental matter multiplets cannot acquire a mass, so they
must always be massless.

The final aspect of the theory that we need to reproduce is the
superpotential. Before the twisting, the $\cN=2$ theory had, in
$\cN=1$ language, two pairs of chiral bi-fundamentals $A_i,B_i$ and
two adjoints $\Phi,\tPhi$, one for each factor of the gauge group. They
were coupled by a standard cubic superpotential as dictated by $\cN=2$
supersymmetry:
\eqn\cubicsup{
W(A_i,B_i,\Phi,\tPhi) = g_1\, \tr\,\Phi\,(A_1B_1 + A_2B_2)
+ g_2\, \tr\,\tPhi\,(B_1A_1 + B_2A_2) }
Now the twisting clearly assigns a mass to the adjoints. The mass
parameter $m$ is actually known\refs\hoo\ to be proportional to
$\mu=\tan\theta$ where $\theta$ is the rotation angle. Because the
model is compactified on $x^6$, it is clear that if the relative
rotation angle between NS 5-brane 1 and 2 is $\theta$, then the angle
between 2 and 1 (going the other way around the circle) is
$-\theta$. Hence the mass parameter is equal and opposite for the two
adjoints, therefore it is antisymmetric under exchange of the two
gauge groups. 

This needs a slight qualification. Physically, the amount of twisting
experienced by 4-branes which are not at the origin will depend
inversely on the separation between the 5-branes, which in turn is
inversely proportional to the square of the gauge
coupling\refs\witfourd. Hence the mass perturbation actually takes the
form 
\eqn\masspert{
W_m(\Phi,\tPhi) = \half m\,\left( (g_1)^2 \tr\,\Phi^2 -
(g_2)^2 \tr\,\tPhi^2 \right) }
Integrating out the adjoints then gives a quartic
superpotential:
\eqn\quarticsup{
W(A_i,B_i) = {1\over 2m}\, \tr\,\left( (A_1B_1A_2B_2) -
 \tr\,(B_1A_1B_2A_2)\right) }
Note that the superpotential is finite as long as
$\theta\ne \pi/2$, it goes smoothly to zero as the branes become
exactly orthogonal\foot{Even though the identification $\mu
=tan\theta$ gives correct answer for many cases, it is actually valid
only for {\it small} $\theta$\refs\ahahan. So for our case at $\theta
=\pi/2$ we can still have some superpotential. We are grateful to
A. Uranga for pointing this out.}.  In what follows, we
will mainly analyse the case of orthogonal branes as it is pictorially
simpler.

This superpotential has a nonabelian global symmetry which becomes
visible if we write it as:
\eqn\symsup{
W(A_i,B_i) = {1\over 2m}\,\epsilon^{ij}\epsilon^{kl}
\tr\,(A_iB_kA_jB_l)}
This has manifest global $SU(2)\times SU(2)$ symmetry, under which the
$A_i$ and the $B_j$ transform separately as doublets under the first
and second factors respectively. We will identify parts of this
symmetry after obtaining the M-theory limit of the model.

\newsec{Brane Configurations and Geometric Singularities in various
Dimensions}

We now give an alternative derivation of the relationship between
branes at conifolds and intersecting brane configurations, which will
help us to see more clearly how the various geometrical data of the
conifold fit together in the brane picture. Although the discussion
will be somewhat qualitative and some constants have to be fixed by
hand, this will provide us a definite map between coordinates in the
brane configuration and in the conifold.

We will proceed in the reverse direction to the previous section, in
the sense that we will start with particular brane configurations and
use U-duality transformations to map them to D3 branes near geometric
singularities.

Let us consider two NS5 branes in type IIA oriented along some directions.
There are three interesting cases:

(i) The 5-branes have all five directions common,

(ii) The 5-branes have three of their directions common, and

(iii)The 5-branes have only one common direction.

The first one gives rise to ALE spaces with $A_{k-1}$ singularity
(here we will focus mainly on $A_1$ singularities, the general case
arises from having more than two NS5 branes). The second case is the
subject of this paper. It gives rise to conifold singularities. The
third case will give rise to toric Hyper-Kahler
manifolds\refs{\gnttown,\lenvafa}. We will see that a certain set
of U-duality transformations relate the various cases.

\vskip.2in

{\it ALE space with $A_{k-1}$ singularity}

\vskip.1in

Let us start with a configuration of a D3-brane suspended between two
parallel D5-branes. The situation can be represented as follows:
$$
\matrix{D5:&~~&1&2&3&4&5&-&-&-&-\cr
D3:&~~&1&2&-&-&-&6&-&-&-\cr}
$$
This will be used, as in Ref.\refs\gutcurio, to obtain the metric for
D4-branes suspended between NS 5-branes.

The first step is easy since it is known\refs\tseytlin\ how to write
down metrics for general configurations of intersecting
D-branes\foot{Consider a system of $N$ intersecting Dp branes with
harmonic functions $H_i$ for each of them. The metric for such a
system follows a general formula. Choose the maximal set of common
directions, say $n_1$, and write the metric for that part with a
factor $(H_1H_2...H_m)^{-1}$. $m$ is the number of D branes which have
$n_1$ common directions.  Now choose the next set.  And so on. In the
end the directions along which no branes lie appear in the metric
without a prefactor. Finally the whole metric is multiplied with
$(H_1H_2H_3.....H_N)^{1/2}$. As an example let $n_1, n_2, n_3$ be the
set of common directions and $m_1,m_2,m_3$ be the number of D branes
with those common directions, then the metric will be
$$
\eqalign{ds^2= &(H_1...H_N)^{1/2}
[(H_1..H_{m_1})^{-1}ds^2_{012..n_1}+ (H_1...H_{m_2})^{-1}ds^2_{n_1+1,..,
n_1+n_2}\cr
&+....+ ds^2_{no~common~directions}]}
$$ 
For reviews, see for example Refs.\refs{\tseytrev,\gnt}.}
For NS-branes, certain
scalings need to be done as we will see.

Let $H_i= 1 + Q_i/ r^d_i$ be the relevant harmonic functions for the
D3 and D5-branes respectively. $Q_i$ is proportional to the charge of
the brane, and $r$ is the transverse distance. The metric for the
above configuration is:
\eqn\metricone{      
\eqalign{ds^2=& (H_5H_3)^{-1/2} ds^2_{012} + (H_3/H_5)^{1/2}ds^2_{345}+
 \cr
&+ (H_5/H_3)^{1/2} ds^2_6
 + (H_3H_5)^{1/2}ds^2_{789}}}
In this case, the $H_i$ are harmonic functions in the three directions
$(x^7,x^8,x^9)$ since these are the ``overall transverse'' directions
in the problem. Hence each of them is of the form $1 + 1/r$ where
$r=((x^7)^2 + (x^8)^2 + (x^9)^2)^\half$. 

Under a S-duality transformation the system becomes a D3 brane between
two NS5 branes.  The metric for this configuration is just the
previous one multiplied by a factor of $H_5^{1/2}$:
\eqn\metrictwo{ 
ds^2 = (H_3)^{-1/2}ds^2_{012} + (H_3)^{1/2}ds^2_{345}+H_5H_3^{-1/2}ds^2_6
+H_5H_3^{1/2}ds^2_{789}}
A T-duality along $x^3$ will now bring the theory to IIA  with a
configuration of a D4 brane between two NS5 branes. The metric for this
configuration will be: 
\eqn\metricthree{
ds^2 = (H_3)^{-1/2}
ds^2_{0123} + (H_3)^{1/2}ds^2_{45} + H_5H_3^{-1/2} ds^2_6 +  H_5H_3^{1/2} 
ds^2_{789}}
At this point we go to IIB via T-duality along $x^6$. The resulting
configuration turns out to be a bunch of D3 branes on a geometric
singularity. This geometry is basically the T-dual manifestation of
the NS5-branes.

The duality relations that we need can be found, for example, in
Ref.\refs\bho. We quote the relevant formulae below ($g$ and $B$ are
the metric and the antisymmetric fields of type IIA and $G$ is the
metric of type IIB, $x^6$ is the compact direction).
\eqn\Gvsg{
G_{mn}= g_{mn}-(g_{6m}g_{6n}-B_{6m}B_{6n})/g_{66},~~~
G_{66}= 1/g_{66},~~~G_{6m} = B_{6m}/g_{66}}
Here $m,n$ take all values from 0 to 9 except 6.  For the IIA metric
in Eq.\metricthree, we have the following metric components:
\eqn\compon{
\eqalign{g_{\mu\nu}=& H_3^{-1/2}\eta_{\mu\nu},~~g_{44}= g_{55}= H_3^{1/2}, 
~~g_{66}= H_5H_3^{-1/2},\cr
& g_{77}= g_{88} = g_{99} = H_5H_3^{1/2}}}

$\mu,\nu = 0, 1, 2, 3$ are the spacetime coordinates.  The
transformation formulae in eq.\Gvsg\ will give the following metric
components for the IIB case:
\eqn\compontwo{
\eqalign{G_{\mu\nu}=& g_{\mu\nu},~~ G_{44}= G_{55}= g_{44}, 
~~G_{66}= (g_{66})^{-1},\cr
& G_{6i}= B_{6i}/g_{66},~~ G_{ii}= g_{ii}+B_{6i}^2/g_{66}}}
$i=7,8,9$ and $B_{6i}= \omega_i$ is the antisymmetric background in
the type IIA picture. $\omega_i$ solves the B-field equation
\eqn\bfield{
\vec{\grad{}} \times\vec{\omega} = \vec{\grad{}} H_5 }
Therefore the IIB metric arising from T-duality
on Eq.\metricthree\ will look like
\eqn\looklike{
ds^2 = G_{\mu\nu}\,ds^2_{0123} + G_{ij}\,ds^2_{ij},~~~~ i,j= 6,7,8,9}
which after putting in all factors becomes: 
\eqn\finmetricone{
ds^2 = 
H_3^{-1/2}ds^2_{0123}+H_3^{1/2}[ds^2_{45}+H_5^{-1}(dx_6+\omega.dx_{789})^2
+H_5ds^2_{789}]}
We observe that the metric describes a D3 brane with world-volume
along (0123)-directions as well as a Taub-NUT space in the transverse
directions (6789). The harmonic functions $H_5$ and $H_3$ are given by
$1+ 1/r$.  Here, however, we face a problem.  $H_5$ (which is still
$1+1/r$) is as desired for the Taub-NUT metric. But $H_3$ should have
been the harmonic function of a 3-brane with 6 noncompact transverse
directions, namely $H_3 = 1+ 1/r^4$ where $r=((x^4)^2 + \ldots
(x^9)^2)$ to get the correct behaviour.  The problem occurs when we
demand a ``localised'' 3-brane, as opposed to the ``delocalised'' one
that occurs in Eq.\metricone. Therefore by following the standard
T-duality rules (and assuming some delocalised directions) we {\it do
not} recover the exact metric of a D3 brane near an $A_{k-1}$
singularity.  A more general analysis wherein no delocalisation is
assumed may lead to a better  result.

{}From Ref.\refs\bho\ it is easy to check that in the IIB case, no other
background will be excited under the above transformations.

At this point, as shown in \refs\gutcurio, we can examine the near
horizon geometry of this system, which turns out to describe type IIB
theory on $AdS_5\times S^5/Z_2$ ($Z_k$ if there are $k$ NS5
branes). To see this we first go near the center of the Taub-NUT
space. For distances very close to this point, we can neglect the
constant part in the $H_5$ harmonic function. This way the D3 brane is
localised at the $A_{k-1}$ singularity. Now in the near horizon region
one can neglect the constant part of $H_3$ also. This way the geometry
resembles $S^5/Z_k$.

The gauge theory on the D4 brane can be read off explicitly from this
model. We have a configuration of $k$ parallel NS5 branes arranged on
a circle $x^6$. The D4 brane is cut at $k$ points to give a gauge
group of $U(1)^k$ (or $U(N)^k$ if there are $N$ D4 branes). The matter
multiplets will come from strings joining two D4 branes across an NS5
brane. These will be hypermultiplets in bi-fundamental
representations. We will return to this later.

\vskip.2in

{\it Conifold singularity}

\vskip.1in

Now we proceed in the same way but for the case relevant to the
$\cN=1$ model that we have presented in the preceding sections.
As before, we start with a configuraton of two orthogonal D5 branes
and a D3 brane between them. The configuration is
$$
\matrix{D5:&~~&1&2&3&4&5&-&-&-&-\cr
D5':&~~&1&2&3&-&-&-&-&8&9\cr
D3:&~~&1&2&-&-&-&6&-&-&-\cr}
$$
Let $H_i$ be the relevant harmonic functions for the D5, D5$'$ and D3
branes. This time, the $H_i$ are harmonic functions in one overall
transverse direction, $x^7$. Hence they behave as $H_i = 1 + r$
where $r = |x^7|$. 

Under a S-duality this will turn into two NS5 branes and a D3
between them. A further T-duality along $x^3$ will give us our
configuration, for the special case where the rotation parameter
$\alpha$ discussed earlier is equal to 1.  This situation is similar
to the previous case of an ALE singularity. Therefore a further
T-duality along $x^6$ should give us a bunch of D3 branes near a
conifold singularity.

The metric for the above configurations of D branes can be written
down following standard prescriptions as explained in the previous
section.  The result is:
\eqn\stdpres{
\eqalign{ds^2=& (H_5H_5'H_3)^{-1/2}ds^2_{012}+
H_3^{1/2}(H_5H_5')^{-1/2}ds^2_3 
+ (H_3H_5')^{1/2}H_5^{-1/2}ds^2_{45}+ \cr
& + (H_3H_5)^{1/2}H_5'^{-1/2}ds^2_{89}
+ (H_5H_5')^{1/2}H_3^{-1/2}ds^2_6 + (H_5H_5'H_3)^{1/2}ds^2_7}}
The metric after a S and a $T_3$ duality will give a configuration of
a D4 brane between two orthogonal NS5 branes. The metric for this
configuration is easy to write down from eq.\stdpres. The result is:
\eqn\metricfour{
\eqalign{ds^2 =& H_3^{-1/2}ds^2_{0123} + H_5'H_3^{1/2} ds^2_{45} 
+H_5H_3^{1/2} ds^2_{89}+ \cr 
&+ H_5 H_5'H_3^{-1/2}ds^2_6 + H_5 H_5'H_3^{1/2}ds^2_7}} 
At this point we use the duality map Eq.\Gvsg\.  On the type IIA side
there can be a nontrivial $B_{\mu\nu}$ background. We assume the non
zero values are $B_{46}, B_{68}$ and $B_{ij}$ for $i,j \in
(4,5,8,9)$. $x^6$ is the compact circle\foot{If we were doing the
reverse procedure, starting with a conifold geometry and using the
T-duality relations to get to our brane picture, then the
configuration would come out as two intersecting branes at an angle to
both 45 and 89 planes. However we can rotate the configuration so that
the two branes are along 45 and 89 respectively, for $\alpha=1$. At a
general value of $\alpha$ we would keep the NS5-brane fixed along 45
and rotate the NS5' brane in the (45,89) planes}. We get the following
metric components for the IIB case:
\eqn\transf{
\eqalign{ G_{\mu\nu}=& g_{\mu\nu},~~ G_{66}= g_{66}^{-1},
~~ G_{6i}= B_{6i}/g_{66},\cr 
&G_{ii}= g_{ii}+B^2_{6i}/g_{66},~~ G_{77}= g_{77},
~~ G_{48}= B_{64}B_{68}/g_{66}}}
where $i=4,8$ and $\mu,\nu=0, 1, 2, 3$ (if we want to get the rotated
case then $i = 4,5,8,9$). The IIB metric therefore becomes:
\eqn\iibbecome{
\eqalign{ds^2=& H_3^{-1/2}ds^2_{0123}+H_3^{1/2}[H_5'ds^2_{45}+H_5ds^2_{89}
+H_5H_5' ds^2_7+ \cr
& + (H_5H_5')^{-1}(ds_6 + B_{64}ds_4+B_{68}ds_8)^2]}} 
In addition to $B_{46},B_{68}$, whose roles will be explained below,
we have chosen to excite constant nonzero B-fields, $B_{ij}$ where
$i,j\in (4,5,8,9)$, on the IIB side\refs\bho. This will be the
nontrivial B background on a conifold. As we will see, this will
parametrise the separation of the NS5-branes, or equivalently the
difference of gauge couplings in the two factors, a quantity that is
not encoded in the geometry of the problem in the type IIB
description.

Now we will argue that the above metric is locally that of a 3-brane
at a conifold. First of all, as in the previous discussion, we must
take $H_3$ to be the harmonic function of a 3-brane localised in 6
transverse directions, while $H_5$ and $H_5'$ continue to be
$(1+|x^7|)$. Now since $H_5,H_5'$ are nonsingular for all finite
$x^7$, we can absorb them into the coordinates $x^4,x^5$ and $x^8,
x^9$. In any case, near the point $|x^7| =0$ we can take the
harmonic functions to be effectively constant.  

Now we see that Eq.\iibbecome\ resembles the form of the metric for a
3-brane at a geometrical conifold singularity. However, in
Eq.\iibbecome, the $(4,5)$ and $(8,9)$ directions are planar, while
for a genuine conifold they need to combine into the direct product of
round spheres with a definite radius. In other words, we need to
make the replacement: 
\eqn\sphere{
ds^2_{45} + ds^2_{89} \to C \sum_{i=1}^{i=2}~(d\theta_i^2 +
sin^2\theta_i\, d\phi_i^2)}
where $C$ is a constant. This amounts to the substitution:
\eqn\amountsto{
dx^4 \to \sqrt{C}\, sin\theta_1\, d\phi_1,\qquad dx^5 \to \sqrt{C}\,
d\theta_1}
and similarly for $(x^8,x^9)$ and $(\theta_2,\phi_2)$. As we will see
again in later sections, the fact that our procedure does not quite
reproduce a conifold, but rather something similar where two 2-spheres
are replaced by 2-planes, is responsible for the fact that only the
$U(1)$ subgroups of the global $SU(2)$ symmetries are manifest.

Next, in analogy with the Taub-NUT case, we define $\omega_4 = B_{64}$
and determine it by solving the equation
\eqn\curlomega{
\vec{\grad{}}\times \vec{\omega} = {\rm constant}}
where $\vec{\omega}=(\omega_4,0,0)$, and the right side is constant
(unlike for the Taub-NUT case) because of the fact that the harmonic
function $H_5$ is linear in $x^7$. In polar coordinates
this becomes:
\eqn\inpolar{
{1\over\sin\theta_1} {\del\over \del\theta_1}(\sin\theta_1\, \omega_4) =
{\rm constant}}
This equation is solved by $\omega_4 = \cot \theta_1$. The same
procedure gives $\omega_8=B_68=\cot\theta_2$. With the replacement
$x^6\rightarrow \psi$, we have:
\eqn\omegasub{
(dx^6 + B_{64} dx^4 + B_{68} dx^8)^2 = 
(d\psi + \cos\theta_1 d\phi_1 + \cos\theta_2 d\phi_2)^2 }
Finally we make the conformal transformation $x^7=\log r$, after which
(suitably rescaling coordinates and making an appropriate choice of
the constant $C$) the term in square brackets in Eq.\iibbecome\
becomes the conifold metric:
\eqn\confinal{
ds_{conifold}^2 = 
dr^2 + r^2\left({1\over 6} 
\sum_{i=1}^2 (d\theta_i^2 + sin^2\theta_i\, d\phi_i^2)  +
{1\over 9}
(d\psi + cos\theta_1 d\phi_1 + cos\theta_2 d\phi_2 )^2\right) }
This metric still solves the supergravity equations of motion, as was
shown in Refs.\refs{\kehagias,\ahm}.

This completes our map from the metric of a configuration of
intersecting branes to that of 3-branes at a conifold (modulo the
heuristic step of replacing two 2-planes by 2-spheres):
\eqn\finmetrictwo{
 ds^2 = (H_3)^{-1/2} ds^2_{0123} + (H_3)^{1/2} [ ds^2_{conifold}]}

We make the following observations:

(i) From eq.\conifold\ we can calculate the base of the conifold by 
intersecting the space of solutions of \conifold\ with a sphere of radius
$r$ in $C^4$,
\eqn\sphere{
\sum^4_{i=1}~| z_i|^2 = r^2}
If we now break up 
$z$ into its real and imaginary parts, $z_i = x_i + iy_i$, then we  have
from \conifold\ and \sphere\
\eqn\base{
x_ix_i = r^2/2, ~~~~ y_iy_i= r^2/2, ~~~~ x_iy_i=0} 
The first of these defines an $S^3$ with radius $r/{\sqrt 2}$. The
other two equations define an $S^2$ bundles over $S^3$. Since all such
bundles are trivial the base has a topology of $S^2\times
S^3$\refs\candelas.

(ii) If the $(4,5)$ and $(8,9)$ directions were planar, then we would
have a $U(1)$ symmetry associated to individual rotations of each of
them. Taking them to be round spheres should enhance these symmetries
to $SU(2)\times SU(2)$. It appears that the enhanced symmetries are
not directly visible in the brane construction, but the above
manipulations (and the discussion to follow) nevertheless illuminate
what they should geometrically correspond to. The coordinates $z_i$
transform as a vector of $SO(4)$, giving rise to a global symmetry
$SO(4) \sim SU(2)\times SU(2)$. Since the $z_i$ also parametrises the
$\cN=1$ chiral multiplets $A_i,B_i;~i=1,2$, these multiplets transform
as a doublet of $SU(2)$. As we will see in the following section, the
lift of this model to M-theory is consistent with the appearance of
this global symmetry, in contrast to the non-elliptic case studied in
Refs.\refs{\hoo,\witneqone,\brand} where these symmetries cannot
appear.

(iii) This mapping of a brane configuration to a geometric manifold
actually alows us to identify locally all the directions of the
conifold in the brane picture. For the previous case of an ALE space
we found that an $S^2$ and the transverse distance $r$ lie in the
(789) direction. In the present case, there are two $S^2$ factors
which play a symmetrical role, these are associated to the (45) and
(89) directions. Therefore, as we have seen above, they are two
supersymmetric cycles in the two NS5 branes. The transverse distance
is again $x^7$. The $U(1)$ fibre of the conifold base is the $x^6$
direction. Finally, the $S^2$ factor in the direct product $S^2\times
S^3$, over which the $U(1)$ part does not vary, is parametrised
locally by two combinations of the coordinates $(x^4,x^5,x^8,x^9)$
that are symmetric under the exchange $(4,5)\leftrightarrow
(8,9)$. The $S^3$ factor is therefore parametrised by the other two
combinations along with $x^6$.

(iv) From Eqs.\finmetrictwo\ and \confinal\ we have:
\eqn\conaga{
ds^2= H_3^{-1/2}ds^2_{0123}+ H_3^{1/2}[dr^2+ G_{ij}ds^2_{ij}]}
where $i,j=4,5,6,8,9$ and $H_3= 1+ L^4/r^4$ with $L^4 = 4\pi g_sN
(\alpha')^2$. Following\refs\klebwit\ we see that the near horizon
limit ($r\to 0$) of the geometry is $AdS_5\times T^{1,1}$.  At this
point we can make the connection between $S^5/Z_2$ and $T^{1,1}$
clear. From Ref.\refs\klebwit\ we expect that when one blows up the
fixed circle of $S^5/Z_2$ one gets the Einstein space $T^{1,1}$. This
simply amounts to rotating the brane configurations from being
parallel to orthogonal.

$AdS_5\times T^{1,1}$ has the explicit form:
\eqn\explicit{
ds^2= (r^2/L^2)ds^2_{0123}+ L^2[dr^2 + G_{ij}ds^2_{ij}]/r^2}
The factor of $L^2$ implies that now the two $S^2$ do not shrink to
zero size. In the brane picture, if the number of D4
branes is very large then in their near-horizon region,
the $S^2$'s have a definite size.

(v) We have seen that a single T-duality along the isometry direction
of a conifold (i.e. the $x^6$ direction) gives us a configuration of
two orthogonal NS5 branes in type IIA theory having a common $3+1$
dimensions.  It is natural to ask how this picture is related to the
one proposed by BSV\refs\bsv. To make this connection we use the fact
that the $S^3$ of the base can be written as a $U(1)$ fibration over
$S^2$. The $U(1)$ and the $S^2$ have already been identified from the
brane picture. 

Now, this $S^2$ can be thought of as a degenerating torus, one of
whose cycles is shrinking to zero size. The point where the cycle
degenerates can be removed and the manifold becomes topologically a
sphere. One can similarly treat the other $S^2$\foot{The two
degenerating tori of Eq.\rewrit\ can be identified with the tori
formed by taking one cycle from each of the $S^2$. This is because
from Ref.\refs\bsv\ we know that the $S^3$ of the base come from $x^6$
and one cycle from each of the two fibre torus.}.  One can T-dualise
along the two directions of the degenerating torus. This will take the
theory back to type IIB and the configuration will be two orthogonal
NS5 branes having a common $3+1$ dimensions. This is the BSV
picture. As we saw earlier, the brane construction of this maps to a
``brane box'' but as far as we can see, it does not provide the
straightforward interpretation of the system that the type IIA picture
gives.

(vi) We can also see the relation between ALE spaces and conifold a
bit more clearly from the brane analysis. For the case of parallel
branes (without the D4 in between) the orthogonal space is ALE$\times
R^6$. As we rotate the branes from the parallel to the orthogonal
configuration, we see that an $R^2$ in $R^6$ starts becoming a
$P^1$. This would imply that the orthogonal space is an ALE fibration
over a $P^1$ which from eq.\conifagain\ is precisely a conifold.

\vskip.2in

{\it Toric Hyper-Kahler manifolds}

\vskip.1in

Although not directly related to the main theme of this paper, we
digress briefly to discuss this case. Here it is more economical to
study the reverse map, namely given a hyper-K\"ahler manifold, we can
use eq.\Gvsg\ to see what kind of brane construction this gives
rise to. We will find that it is a set of intersecting NS5-branes
along a string. This part will be mainly a review of
Refs.\refs{\gnttown,\lenvafa}.

The toric eightfold that we are interested in will have $T^2$
isometry. A generic toric eightfold has the following local form of
the metric:
\eqn\toric{ 
ds^2 = U_{ij}dx^idx^j + U^{ij}(d\phi_i+A_i)(d\phi_j+A_j)}
where $U_{ij}$ are the entries of a positive definite symmetric 
$n\times n$
matrix function $U$ of the $n$ set of coordinates $x^i$. And $A_i =
dx^j\omega_{ji}$ where $\omega$ is a $n\times n$ matrix. $\phi_i$ are the 
two isometry directions.

Therefore we have a configuration of type IIA theory on a eightfold
with a metric
\eqn\metricfive{
ds^2 = ds^2_{01}+ U_{ij}dx^idx^j + U^{ij}(d\phi_i+A_i)(d\phi_j+A_j)} 
If we T-dualise twice along the two isometry directions using the
relations given in Eq.\Gvsg\ we get a model back in IIA which can be
interpreted as an arbitrary number of NS5-branes intersecting on a
string\refs\gnttown\ (for subsequent developments, see
Refs.\refs\papad).

Starting with M-theory on the toric fourfold, we compactify along one
isometry direction and then T-dualise along other, which takes us to
type IIB theory with a configuration of one NS5-brane and a D5
brane having $2+1$ common directions. A 2-brane probe in M-theory
will now become a D3-brane suspended between the two 5-branes. This
picture, as argued in\refs\gnttown, is not the Hanany-Witten model
because we get $\cN=3$ in $d=3$ from this model. A generalisation of
this model and other questions related to the near horizon geometry
etc. will be addressed in a future paper\refs\wkinprogress.

\newsec{The M-theory Description}

Remarkable insight can be gained into the dynamics of theories
constructed via type IIA branes by taking the strong-coupling limit
and going to M-theory. This approach was pioneered by Witten in
Ref.\refs\witfourd\ for $\cN=2$ theories, where it yields an amazing
new derivation of the Seiberg-Witten curves and their generalizations
for various gauge groups, and also gives rise to the solution found in
Ref.\refs\donagiwitten\ of the conformal $\cN=2$ models related to
integrable systems. It was generalised to $N=1$ models in
Refs.\refs{\hoo,\witneqone,\brand}, and subsequently studied
by many other authors.

The common feature of these solutions is that a configuration of
D4-branes ending on NS5-branes in type IIA string theory has to turn
into a configuration purely made up of 5-branes in the M-theory limit,
since there are no 4-branes in M-theory. Indeed, since M5-branes
cannot actually ``end'' on each other because of charge conservation,
so what actually happens is that there is a single smooth M5-brane at
the end, whose weak-coupling (small M-circle radius) limit looks like
branes ending on branes. On the other hand, the gauge theory that is
being realised via brane configurations has coupling constants that
depend only on the ratio of brane separations to the string coupling,
so by scaling both of these up together, we can keep the gauge theory
coupling finite and still justify the use of M-theory.

Clearly it is an interesting problem to understand what is the
solution of our model using this M-theory limit.  For clarity, let us
list the four different models that we will be comparing in our
discussion:

(i) Non-elliptic $\cN=2$: A model of 2 parallel NS5-branes with $N$
D4-branes stretched between them (where the NS5-branes fill the
directions $(x^1,x^2,x^3,x^4,x^5)$ and are separated along $x^6$,
while the D4-branes fill $(x^1,x^2,x^3)$ and are stretched along
$x^6$, see Fig. 2),
\bigskip

\centerline{\epsfbox{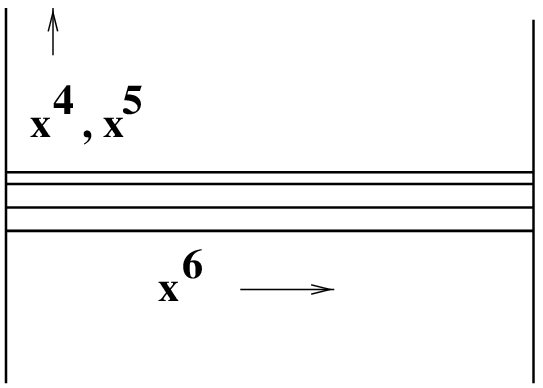}}\nobreak
\centerline{Fig. 2: Model (i): Non-elliptic $\cN=2$}
\bigskip

(ii) Elliptic $\cN=2$: A model of 2 parallel NS5-branes located on a
compact direction, with $N$ D4-branes stretched between them from both
sides (the directions are precisely as for model (i) but $x^6$ is
compact, see Fig. 3),
\bigskip

\centerline{\epsfbox{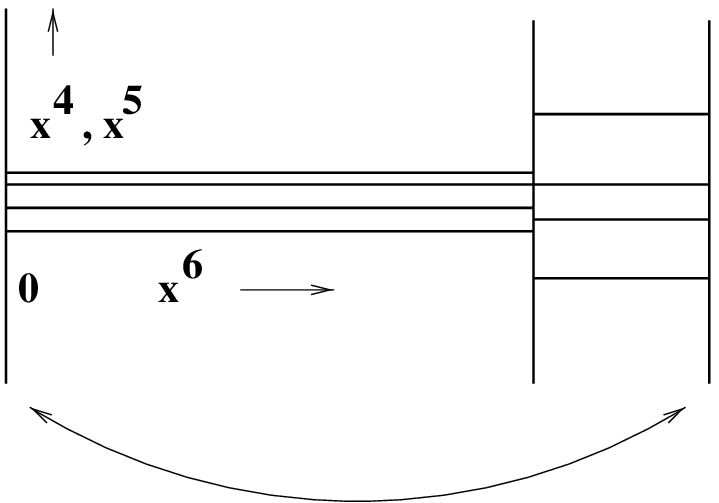}}\nobreak
\centerline{Fig. 3: Model (ii): Elliptic $\cN=2$}
\bigskip

(iii) Non-elliptic $\cN=1$: A model of 2 orthogonal NS5-branes with
$N$ D4-branes stretched between them (where first NS5-branes fills the
directions $(x^1,x^2,x^3,x^4,x^5)$ and is separated along $x^6$ from
the second NS5'-brane, which fills $(x^1,x^2,x^3,x^8,x^9)$, while the
D4-branes fill $(x^1,x^2,x^3)$ and are stretched along $x^6$, see
Fig. 4),
\bigskip

\centerline{\epsfbox{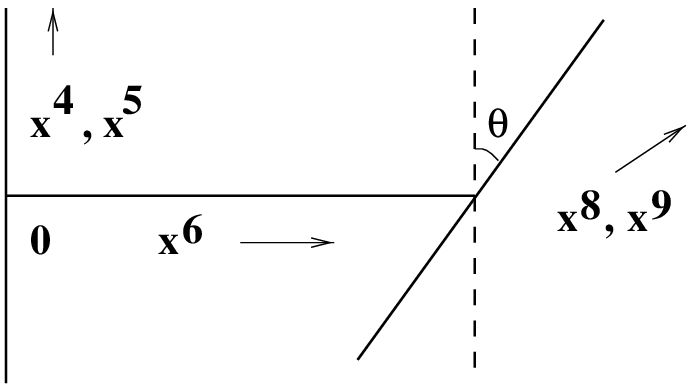}}\nobreak
\centerline{Fig. 4: Model (iii): Non-elliptic $\cN=1$}
\bigskip

(iv) Elliptic $\cN=1$: A model of 2 orthogonal NS5-branes located on a
compact direction, with $N$ D4-branes stretched between them from both
sides (the directions are precisely as for model (iii) but $x^6$ is
compact, this model has already been illustrated in Fig. 1).

Models (i) and (iii) are not conformally invariant -- they can be
thought of as the $\cN=2$ and $\cN=1$ supersymmetric versions of pure
$SU(N)$ QCD. The elliptic models (ii) and (iv) are the ones dual to
3-branes at a $Z_2$ ALE singularity and a conifold, respectively, that
have been the subject of previous sections.

Models (i) and (ii) were solved in the M-theory limit in
Ref.\witfourd. The solutions can be summarised as follows: in model
(i), the M-theory brane is wrapped on $R^4\times\Sigma$ where $R^4$ is
described by the coordinates $(x^0,x^1,x^2,x^3)$ and $\Sigma$ is a
complex curve (Riemann surface) holomorphically embedded in $R^3\times
S^1$ parametrised by the complex coordinates $(v,t)$ where
$v=x^4+ix^5$ and $t=\exp(-(x^6 + i x^{10}))$. The curve $\Sigma$ is
given by the equation
\eqn\neqtwocurve{
t^2 + t(v^N + u_2 v^{N-2} + \ldots + u_N) + 1 = 0}
This curve has $(N-1)$ complex parameters. This fits with the fact
that it has Euler characteristic $\chi=4 - 2N$ which, via $\chi=2 -2g$,
determines the genus $g$ to be $N-1$. The Euler characteristic is
computed by noting that within each of the NS5-branes we have a
2-sphere or $P^1$ (in the $(4,5)$ directions) with $N$ little tubes
joining the two $P^1$'s. Each $P^1$ has $\chi=2$ and each tube takes
away 2 from $\chi$. The situation is illustrated in Fig. 5.
\bigskip

\centerline{\epsfbox{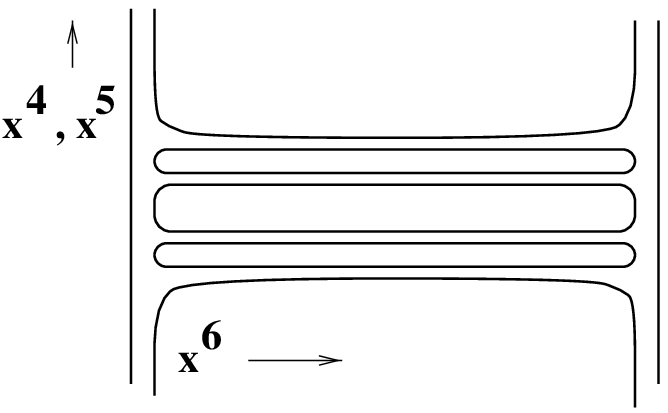}}\nobreak
\centerline{Fig. 5: Solution of model (i): genus = $N-1$}
\bigskip

The genus of the curve is the number of massless photons in the
Coulomb branch of the field theory. We can check this by noting that
in the present case, the gauge group is $SU(N)$ which has a Coulomb
branch of dimension $N-1$.

Model (ii) requires a slightly different approach. This time the
spectrum is a $U(1)\times SU(N)\times SU(N)$ gauge theory with 2
bi-fundamental hypermultiplets (the $U(1)$ is decoupled). The theory
is conformally invariant. The M5-brane to which the brane
configuration tends in the limit of large type IIA coupling is again
wrapped on $R^4\times
\Sigma$ where $R^4$ is the same as for model (i), but now $\Sigma$ is
a complex curve holomorphically embedded in $R^2\times T^2$. The torus
$T^2$ is described by the coordinates $x^6,x^{10}$, both of which are
compact.

If the $(6,10)$ torus is parametrised by two complex coordinates
$(x,y)$ satisfying a Weierstrass equation
\eqn\weier{
y^2 = x^3 + f x + g}
then the curve $\Sigma$ is described by the equation
\eqn\ellipcurve{
v^N + \sum_{i=1}^N f_i(x,y) v^{N-i} =0}
where $f_i(x,y)$ are $N$ meromorphic functions, each of which has a
simple pole at two points $(X,Y)$ and $(X',Y')$ (with equal and
opposite residues). 

Each of the functions depends on two complex parameters, one being the
common residue at the poles and the other a constant shift. Thus we
have $2N$ real parameters, but one of them describes the masses of the
two hypermultiplets, which are equal and opposite. This parameter has
been identified in Ref.\refs\witfourd to be the residue of $f_1$. The
remaining $2N-1$ parameters describe the Coulomb branch of $U(1)\times
SU(N)\times SU(N)$. The parameter for the decoupled $U(1)$ has also
been identified -- it is a constant shift in $f_1$\refs\witfourd.

It follows that the genus of the curve $\Sigma$ should be $2N-1$,
which can be confirmed by noticing that we now have two $P^1$'s (one
on each NS5-brane) joined by $N$ thin tubes from {\it both} sides
along a circle, as in Fig. 6.
\bigskip

\centerline{\epsfbox{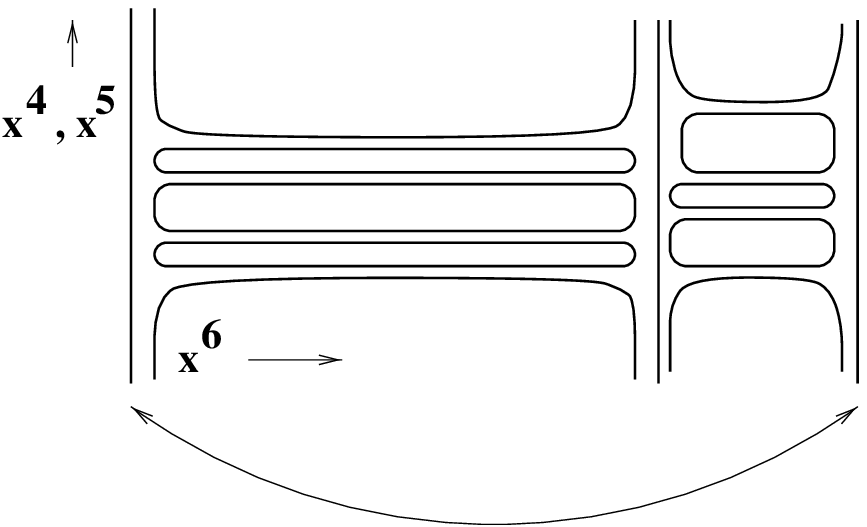}}\nobreak
\centerline{Fig. 6: Solution of model (ii): genus = $2N-1$}
\bigskip

Model (iii) has $\cN=1$ supersymmetry. The corresponding field theory
is usually known as SQCD. The solution of this model can be found in
Refs.\refs{\hoo,\witneqone}. Although it can be obtained
directly, a very useful way to understand the solution is via a
``rotation'' of model (i). The second NS5-brane of model (i) is
rotated in the $(4,5),(8,9)$ hyperplane until it fills the $(x^8,x^9)$
directions (the $w$-plane, where $w=x^8+ix^9$) and lies at a point in
$(x^4,x^5)$, say $v=0$. The effect of this rotation, as we have
discussed for the analogous elliptic model in previous sections, is to
induce a mass term for the adjoint, breaking $\cN=2$ supersymmetry to
$\cN=1$.

It turns out that one cannot perform this rotation at any arbitrary
point of moduli space. This is the brane manifestation of a well-known
phenomenon first noted in Ref.\refs\seiwitone\ for $SU(2)$ gauge
theory, that an adjoint mass term can be turned on (partially breaking
supersymmetry and leading to confinement) only at points on the
$u$-plane where the Seiberg-Witten torus has a node, in other words
where it degenerates to genus 0. In the present context, the D4-branes
lying between the NS5-branes must first come together at the origin,
or else they will get twisted by the rotation and all supersymmetry
will be broken. (The origin is of course replaced non-perturbatively
by a pair of points, where monopoles and dyons become massless, hence
there are actually two points in moduli space where rotation is
allowed.) In short, rotation of an NS5-brane breaking $\cN=2$ to
$\cN=1$ is possible only when the curve $\Sigma$ has degenerated to
genus 0.

The resulting model, in the M-theory limit, is described by an
M5-brane wrapped over $R^4\times\Sigma$ where again $R^4$ is spanned
by $(x^0,x^1,x^2,x^3)$ but this time $\Sigma$ is a certain complex
curve holomorphically embedded in $R^4\times T^2$. Thus in the $\cN=1$
models, the curve is a holomorphic embedding in a non-compact complex
Calabi-Yau 3-fold rather than a complex surface (2-fold) as was the
case for $\cN=2$.

The 3-fold is parametrised by the complex coordinates $v,w$ (for the
$R^4$ part which is made up of the $(4,5)$ and $(8,9)$ directions) and
$(x,y)$ which describe the 2-torus $T^2$. The fact that $\Sigma$ in
this model has genus 0 is evident both from the simple fact that the
Coulomb branch has disappeared, and from the brane construction in
which the two $P^1$'s (one in each 5-brane) are connected by a single
``tube'' consisting of all the 4-branes bunched together. The
situation is depicted in Fig. 7.
\bigskip

\centerline{\epsfbox{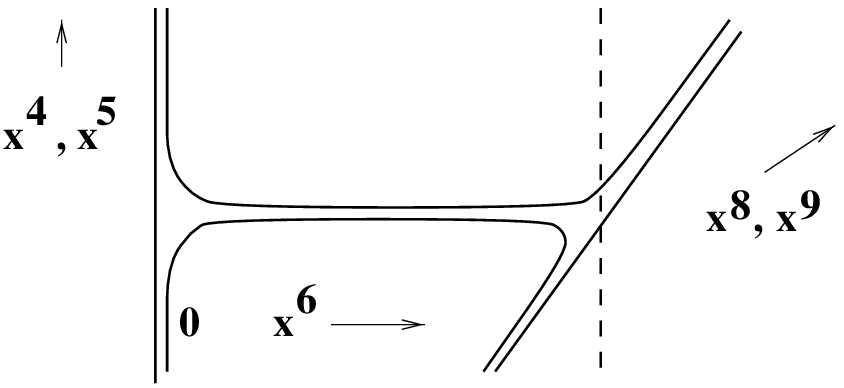}}\nobreak
\centerline{Fig. 7: Solution of model (iii): genus = 0}
\bigskip

Now let us come to model (iv), the one which is the main subject of
this paper. We can try to see it as a rotation of model (ii). Indeed,
this rotation is just the relevant perturbation which in geometric
language takes D3-branes transverse to ALE$\times R^2$ to D3-branes at
a conifold. Let us ask now what is the condition for the model (ii) to
admit a rotation breaking $\cN=2$ supersymmetry to $\cN=1$. In this
case it is easy to see that the Seiberg-Witten curve only has to
degenerate to genus 1 to permit the rotation. However, there is
another constraint, that the hypermultiplet mass parameter must also
go to zero. This is shown as follows.

Geometrically, in order to rotate one 5-brane we need that the
4-branes connecting the 5-branes (from both sides) coalesce
completely. This correponds to going to the origin of the Coulomb
branch for the $SU(N)\times SU(N)$ part of the gauge group (this
collapses each of the two bunches of 4-branes), and also making the
single hypermultiplet mass parameter zero (this makes the two bunches
of 4-branes coincide where they touch the 5-branes). This process
freezes altogether $2N-1$ parameters, leaving only the constant shift
in $f_1$. as a result, the curve $\Sigma$ relevant to this model has
genus 1.

This is of course confirmed by the fact that the Coulomb branch of
this model just corresponds to the free decoupled $U(1)$. Moreover,
geometrically, we have again $N$ D4-branes joining the NS5-branes from
both sides, as in model (ii), but now that they are all collapsed,
there is only a single tube connecting a pair of $P^1$'s from both
sides, with the result again that the genus is equal to 1.

Now we can ask, in the spirit of
Refs.\refs{\witfourd,\hoo,\witneqone}, what is the description of the
genus 1 curve $\Sigma$ on which the M 5-brane is wrapped. The answer
is rather remarkable and is related to a phenomenon recently discussed
by Dorey and Tong in Ref.\refs\tong\foot{We are grateful to David Tong
for explaining these results to us before publication.}. The situation
considered in Ref.\refs\tong\ is a pair of NS5-branes with some D4-branes
stretched between them, and with additional semi-infinite 4-branes on
either side. The particular configuration of interest arises when some
of D4-branes ``line up'' on opposite sides of the NS5-brane (Fig. 8).
\bigskip

\centerline{\epsfbox{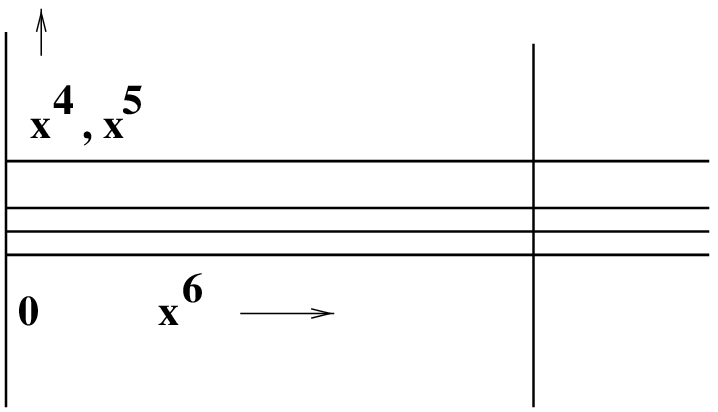}}\nobreak
\centerline{Fig. 8: 4-branes ``lined up'' across a 5-brane}
\bigskip

One can ask what describes the M-theory limit of this configuration,
which correponds to special regions of moduli space. The answer turns
out to be that while the NS5-brane turns into an M5-brane as usual,
the D4-branes that are lined up on opposite sides of it turn into
M5-branes that {\it go through} it. In other words, in the M-theory
limit, the charge carried by the type IIA D4-branes does not flow onto
the 5-brane on which they end from both sides, but goes right
through. The situation is illustrated in Fig. 9.
\bigskip

\centerline{\epsfbox{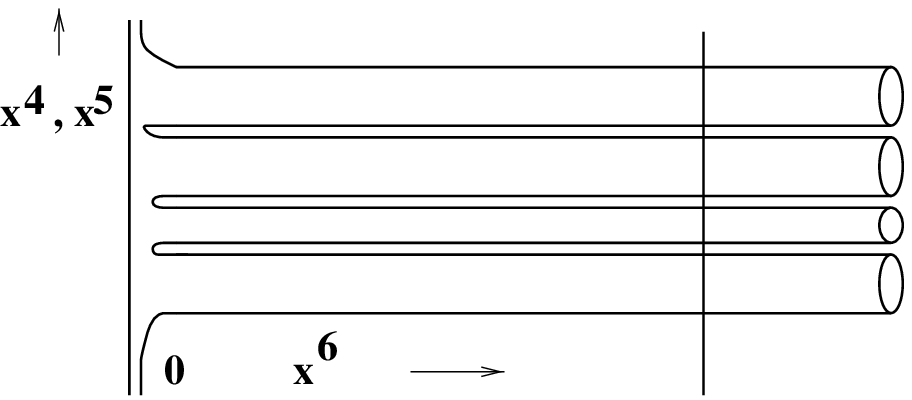}}\nobreak
\centerline{Fig. 9: Cylindrical M5-branes going through transverse
M5-branes}  
\bigskip

In our model, called model (iv) above, this situation is indeed
realised. As explained above, in order to allow rotation of an
NS5-brane and break supersymmetry from $\cN=2$ to $\cN=1$, the
D4-branes of model (ii) must all collect at the origin. Hence by the
above argument, in the M-theory limit we will have a bunch of $N$
coincident toroidal M5-branes wrapped on the $(x^6,x^{10})$, going
through a pair of M5-branes: one aligned in the $(x^4,x^5)$ direction
and the other rotated in the $(x^4,x^5),(x^8,x^9)$ hyperplane (Fig. 10).
\bigskip

\centerline{\epsfbox{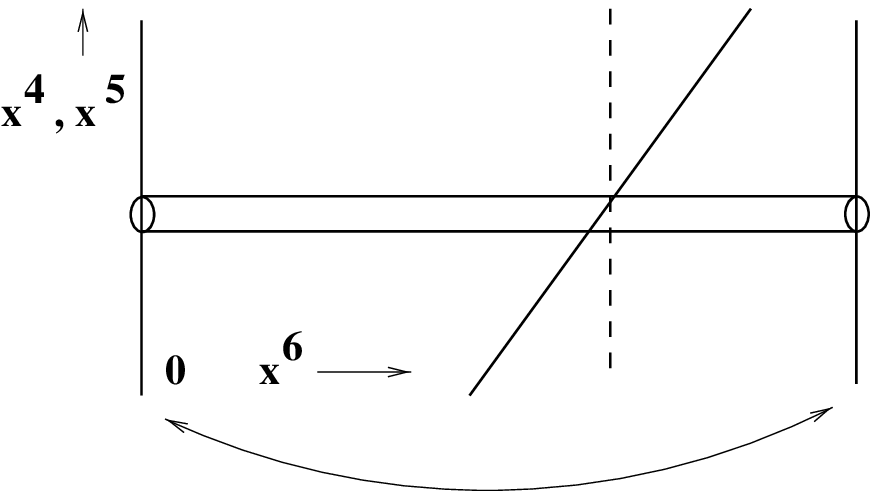}}\nobreak
\centerline{Fig. 10: Solution of model (iv): genus = 1}
\bigskip

This can be seen starting from the M-theory curve for model
(ii). Recall that this model is described by Eq.\ellipcurve, specified
by a set of $N$ complex meromorphic functions $f_i(x,y)$ on the
$(x^6,x^{10})$ torus. We want to consider the special point in moduli
space where we are at the origin of the Coulomb branch and the
hypermultiplet mass parameter is zero. From the discussion above, it
follows that $f_1(x,y)=0$ (the constant part is zero since the
centre-of-mass $U(1)$ parameter is fixed at the origin, and the
residue vanishes because the hypermultiplet mass is zero). Similarly,
since the constant and residue parts of the other $f_i(x,y)$ describe
the VEV's associated to going to the Coulomb branch of the two $SU(N)$
factors of the gauge group (before rotation), to go to the origin of
the Coulomb branch these must also be set to zero, as a result of
which all the $f_i$'s vanish identically. Hence we get the curve
\eqn\neqtwoellip{
v^N=0}
corresponding to $N$ coincident M5-branes wrapped on the
$(x^6,x^{10})$ torus.  After rotation, the curve is no longer given by
a single equation in $R^2\times T^2$ but rather by two equations in
$R^4\times T^2$. By the symmetry that was present between the two
NS5-branes before rotation, the equations must be
\eqn\neqoneellip{
\eqalign{v^N &=0\cr
w^N &=0\cr}}
where $w$ labels the coordinate on the rotated brane. If the rotation
is by 90 degrees then $w=x^8 + ix^9$, otherwise it is a suitable
linear combination of the $(x^4,x^5)$ and $(x^8,x^9)$ coordinates. 

These equations precisely describe the $N$ toroidal branes, which are
located at the origin of the $v$ and $w$ planes. One has to supplement
these by hand with the requirement that the transverse planar branes
are still present, at two points $p_1,p_2$ on the $(x^6,x^{10})$
torus. Hence the M-theory limit has three separate and disconnected
5-brane components, two of which are the original NS and NS' 5-branes
and the third is the torus which has $N$ coincident M5-branes wrapped
on it, described by Eq.\neqoneellip.

The above description has important implications for the global
symmetries and other properties of the model, as we will now see.

\newsec{Global Symmetries and Moduli}

The fact that the stretched branes are decoupled from the transverse
ones makes it clear that rotations of the $v$-plane and $w$-plane can
be carried out independently of each other and of shifts in the $x^6$
and $x^{10}$ directions. This is in contrast to model (iii), where the
the $v$ and $w$-plane rotations and the rotation along $x^{10}$ were
linked together into a single $U(1)$ symmetry by the constraint that
all the branes merge into a genus-0 curve\refs\witneqone. We will
identify some of these transformations as factors of the global
$SU(2)\times SU(2)$ symmetries associated to the conifold and
discussed at length in Ref.\refs\klebwit, and another one as the
$U(1)$ R-symmetry.

As we already briefly pointed out, the $U(1)$'s rotating the $v$ and
$w$ planes will act on the chiral multiplets coming from ``short
strings'' connecting the 4-branes across a 5-brane. Across the first
5-brane, we will get a pair of chiral fields, one in the $(N,\bN)$
representation of the gauge group and the other in the $(\bN,N)$. Let
us label these as $A_1$ and $B_2$. Similarly the chiral fields arising
across the second 5-brane are $B_1,A_2$, transforming in the $(\bN,N)$
and $(N,\bN)$ representation respectively.

Now rotation of the $4,5$ plane by an angle $\alpha$ induces a
transformation on the chiral fields as follows. The short strings
living on the D4-brane would give rise to an $\cN=4$ multiplet if no
NS5-branes were present. The NS5-branes break this to a
hypermultiplet. From the geometry of the problem, it is clear that
this hypermultiplet is the one which parametrised motion of the
4-brane in the $(x^4,x^5,x^8,x^9)$ directions. It decomposes
after breaking to $\cN=1$ (which is induced by a different NS5-brane
located some distance away) into the two chiral multiplets $A_1$ and
$B_2$, hence these two fields can be thought of as being associated to
the $(x^4,x^5)$ and $(x^8,x^9)$ directions respectively. It follows
that under rotation of the $(x^4,x^5)$ plane by an angle $\alpha$,
$A_1$ picks up a phase $e^{i\alpha}$ while $B_2$ is unchanged. On the
NS5'-brane, the situation is reversed. There the chiral multiplets are
$B_1$ and $A_2$, but this brane is also rotated at 90 degrees to the
previous one, hence we find that $A_2$ is charged under $(x^4,x^5)$
rotations. Moreover, because $A_2$ comes from a ``left-pointing''
short string (this is clear since it is an $(N,\bN)$ of the gauge
group), it picks up a phase $e^{-i\alpha}$ under this rotation.

Combining, we find that under rotations of the $(x^4,x^5)$ plane, the
chiral fields transform as
\eqn\fourfiverot{
A_1 \to e^{i\alpha} A_1,\quad A_2\to e^{-i\alpha} A_2,\quad B_1\to
B_1,\quad B_2\to B_2}
This shows that these rotations define the $U(1)$ Cartan subalgebra of
the global $SU(2)$ that is expected in this field theory. This in turn
lends support to our earlier observation that $SU(2)$ arises by
``enhancement'' of this $U(1)$ rotation, if the $(x^4,x^5)$ directions
are compactified on a round 2-sphere.

Similar considerations can be used to show that rotations of the
$(x^8,x^9)$ plane generate a $U(1)$ lying in the second $SU(2)$.

Finally, a shift along $x^6$ generates an identical phase on
all the chiral fields.
\eqn\unonesix{
A_1 \to e^{i\gamma}A_1,\quad A_2\to e^{i\gamma} A_2,\quad
B_1\to e^{i\gamma} B_1,\quad B_2\to e^{i\gamma}B_2 }
This must therefore be the $U(1)$ R-symmetry of the theory. Notice
that we have already identified $x^6$ with the conifold coordinate
that is usually called $\psi$, whose shift precisely generates the
R-symmetry. 

Next, let us consider the discrete symmetries. According to
Ref.\refs\klebwit, there is supposed to be a global $Z_2$ symmetry
that acts on the conifold as $z_4\rightarrow -z_4$ and, in field theory
language, exchanges the two $SU(N)$ factors in the gauge group. Hence
it also exchanges the chiral fields $A_i$ and $B_i$. This symmetry
emerges very neatly from our construction. Recall that in the
beginning of the argument, we had treated the conifold as a fibration
of ALE spaces over a base, which was chosen to be parametrised by
$z_4$. The result of various dualities gave a pair of intersecting
NS5-branes where the base $z_4$ was equal to the sum and difference of
the brane coordinates. Thus, inversion of $z_4$ is nothing but
exchange of the two NS5-branes. This obviously exchanges the factors
of the gauge group and the chiral multiplets associated to each
5-brane.

Another interesting discrete symmetry is the element $w$ of the centre
of SL(2,Z) in the type IIB description:
\eqn\elw{
w = \pmatrix{-1&0\cr 0&-1\cr}}
This was argued in Ref.\klebwit\ to exchange the two factors of the
gauge group, while not exchanging $A_i$ and $B_i$, but instead
exchanging the fundamental representation $N$ of the first $SU(N)$
with the anti-fundamental $\bN$ of the second, and vice-versa. In our
picture this symmetry is just the geometrical inversion of both the
torus directions:
\eqn\refltorus{
x^6\rightarrow -x^6,\qquad\quad x^{10}\rightarrow - x^{10}}
which clearly performs the required tasks. 

Finally, we turn to the moduli. The M-theory solution of our model has
a set of $N$ M5-branes wrapped on a 2-torus parametrised by
$x^6,x^{10}$. We will now examine this point more closely. As is
well-known, type IIB theory arises from M-theory by compactifying on a
2-torus and performing T-duality along one of the cycles. In the
present case, $x^6$ is precisely the direction along which we
performed T-duality to obtain our model from its type IIB description
as 3-branes at a conifold. Hence this torus holds the key to
understanding the relation to type IIB.

In the type IIA theory, the two NS5-branes are located at definite
values of $x^6$, which we may call 0 and $a$. The gauge coupling
constants $g_1$ and $g_2$ of the two $SU(N)$ factors are then
determined, by standard arguments\refs\witfourd\ to be:
\eqn\coupconst{
\eqalign{
{1\over (g_1)^2} &= {R_6 - a\over g_{st}}\cr
{1\over (g_2)^2} &= {a\over g_{st}}\cr}}
This can be recast in a useful form as
\eqn\recast{
\eqalign{
{1\over (g_1)^2} + {1\over (g_2)^2} &= {R_6\over g_{st}}\cr
{1\over (g_1)^2} - {1\over (g_2)^2} &= {R_6 - 2a\over g_{st}}\cr}}
It follows that the difference in couplings is determined by $a$, the
brane separation, while the sum of the couplings is determined only by
the string coupling $g_{st}$ (for fixed $R_6$).

In the M-theory limit, the two planar M5-branes are located at
definite points on the $(x^6,x^{10})$ torus. Hence we have two complex
parameters. One of these generalises $a$, and we will call it $\phi$,
while the other generalises ${1\over g_{st}}$ and is in fact the
complex structure parameter of the 2-torus, which in IIB becomes the
dilaton-axion parameter $\tau_{IIB}$. At the same time on the field
theory side, $g_1$ and $g_2$ are part of complex parameters (by
incorporating the $\theta$-angles) which we will call $\tau_1^{YM}$
and $\tau_2^{YM}$. The obvious holomorphic generalization of the above
equations to the M-theory limit is:
\eqn\holom{
\eqalign{
\tau_1^{YM} + \tau_2^{YM} &\sim \tau_{IIB}\cr
\tau_1^{YM} - \tau_2^{YM} &\sim {\phi \tau_{IIB}}\cr}}
It follows that the sum of the gauge theory parameters is determined
by the (complex) type IIB string coupling, while the difference is
determined by $\phi$. This is identical to relations derived in Sec.3
of Ref.\klebwit\ (where they were derived using very different
reasoning) if we make the identification:
\eqn\ident{
\phi = 2\int_{S^2}(B_{NS}+ i B_{RR}) -1}
where the $S^2$ is one of the factors in the conifold base $S^2\times
S^3$. Here, $B_{NS}$ and $B_{RR}$ are the two 2-form fields in the
type IIB theory, and their appearance in this context is related to
considerations in Ref.\refs\lnv.

This remarkable agreement moreover clarifies the meaning of the $x^6$
separation between the branes in conifold language, as was promised in
Sec. 5 when we allowed constant B-fluxes in the $(4,5,8,9)$
directions. Indeed, we now see the more general result that $\phi$,
the complex distance between the two planar 5-branes, is determined by
the value of the NS-NS and RR B-fields in the dual type IIB theory.
Its real part $\int_{S^2} B_{NS}$ is the separation in
$x^6$. Apparently $\int B_{NS}=\half$ corresponds to 5-branes
located symmetrically at opposite points of the 2-torus, hence
identical couplings and theta angles in the two gauge factors.

In the type IIA description, $B_{NS}$ remains a 2-form while $B_{RR}$
turns into the RR 3-form with one index along the 6 direction (we
T-dualised along a direction orthogonal to the $S^2$, namely $x^6$).
In the M-theory limit, $B_{NS}$ also becomes the M-theory 3-form, with
one index in the ``10'' direction. So finally, in the M-theory picture
there are two constant 3-form backgrounds, with the 3-form having one
index in a toroidal direction and the other two indices outside. This
is of course the familiar way in which the two B-fields of type IIB
originate from M-theory.

\bigskip\bigskip

\noindent{\bf Acknowledgements:} We are especially grateful to Angel 
Uranga for detailed discussions and for sharing the results of
Ref.\refs\angel\ before publication, and to Shishir Sinha for
intensive discussions on aspects of this work. We are also happy to
acknowledge useful conversations with Gottfried Curio, Atish
Dabholkar, Debashis Ghoshal, Steve Gubser, Dileep Jatkar, Albion
Lawrence, Shiraz Minwalla, Matt Strassler, David Tong, Cumrun Vafa and
Edward Witten.  The work of K.D. was supported in part by DOE grant
No. DE-FG02-90ER40542. S.M. acknowledges financial support and
hospitality from Andy Strominger and the Physics Department at Harvard
University, where part of this work was performed.

\listrefs    
\end